\documentclass[12pt]{article}
\usepackage{amsmath,amssymb,latexsym,epsf,epsfig,graphicx,tikz}

\begin{document}

\newcommand{\sect}[1]{\setcounter{equation}{0}\section{#1}}
\renewcommand{\theequation}{\thesection.\arabic{equation}}
\newcommand{\prt}{\partial}
\newcommand{\II}{\mbox{${\mathbb I}$}}
\newcommand{\CC}{\mbox{${\mathbb C}$}}
\newcommand{\RR}{\mbox{${\mathbb R}$}}
\newcommand{\QQ}{\mbox{${\mathbb Q}$}}
\newcommand{\ZZ}{\mbox{${\mathbb Z}$}}
\newcommand{\NN}{\mbox{${\mathbb N}$}}
\def\G{\mathbb G}
\def\UU{\mathbb U}
\def\S{\mathbb S}
\def\T{\mathbb T}
\def\tS{\widetilde{\mathbb S}}
\def\V{\mathbb V}
\def\tV{\widetilde{\mathbb V}}
\newcommand{\D}{{\cal D}}
\def\hint{H_{\rm int}}
\def\R{{\cal R}}

\newcommand{\rd}{{\rm d}}
\newcommand{\diag}{{\rm diag}}
\newcommand{\U}{{\cal U}}
\newcommand{\K}{{\mathcal K}}
\newcommand{\cP}{{\cal P}}
\newcommand{\dQ}{{\dot Q}}
\newcommand{\dS}{{\dot S}}

\newcommand{\pnf}{P^N_{\rm f}}
\newcommand{\pnb}{P^N_{\rm b}}
\newcommand{\hnf}{P^Q_{\rm f}}
\newcommand{\hnb}{P^Q_{\rm b}}

\newcommand{\ph}{\varphi}
\newcommand{\phd}{\widetilde{\varphi}} 
\newcommand{\psis}{\psi_{\rm s}}
\newcommand{\psib}{\psi_{\rm b}}
\newcommand{\psds}{\widetilde{\varphi}^{(s)}}
\newcommand{\phdb}{\widetilde{\varphi}^{(b)}}
\newcommand{\lambdad}{\widetilde{\lambda}}
\newcommand{\tx}{\widetilde{x}} 
\newcommand{\etat}{\widetilde{\eta}}
\newcommand{\phl}{\varphi_{i,L}}
\newcommand{\phr}{\varphi_{i,R}}
\newcommand{\phz}{\varphi_{i,Z}}
\newcommand{\mum}{\mu_{{}_-}}
\newcommand{\mup}{\mu_{{}_+}}
\newcommand{\mupm}{\mu_{{}_\pm}}
\newcommand{\muv}{\mu_{{}_V}}
\newcommand{\mua}{\mu_{{}_A}}
\newcommand{\mut}{\widetilde{\mu}}

\def\a{\alpha}
 
\def\A{\mathcal A} 
\def\H{\mathcal H} 
\def\U{\mathcal U} 
\def\E{\mathcal E} 
\def\C{\mathcal C} 
\def\L{\mathcal L} 
\def\O{\mathcal O}
\def\I{\mathcal I}
\def\der{\partial }
\def\mis{{\frac{\rd k}{2\pi} }}
\def\ri{{\rm i}}
\def\xt{{\widetilde x}}
\def\ft{{\widetilde f}}
\def\gt{{\widetilde g}}
\def\qt{{\widetilde q}}
\def\tt{{\widetilde t}}
\def\tmu{{\widetilde \mu}}
\def\prt{{\partial}}
\def\tr{{\rm Tr}}
\def\inc{{\rm in}}
\def\out{{\rm out}}
\def\li{{\rm Li}}
\def\e{{\rm e}}
\def\eps{\varepsilon}
\def\k{\kappa}
\def\v{{\bf v}}
\def\ebf{{\bf e}}
\def\abf{{\bf A}}
\def\fa{{\mathfrak a}} 


\newcommand{\finprf}{\null \hfill {\rule{5pt}{5pt}}\\[2.1ex]\indent}

\pagestyle{empty}
\rightline{September 2016}

\bigskip 

\begin{center}
{\Large\bf Quantum Transport in Presence of Bound States - Noise Power}
\\[2.1em]

\bigskip

{\large Mihail Mintchev}\\ 
\medskip 
{\it  
Istituto Nazionale di Fisica Nucleare and Dipartimento di Fisica, Universit\`a di
Pisa, Largo Pontecorvo 3, 56127 Pisa, Italy}
\bigskip 

{\large Luca Santoni}\\ 
\medskip 
{\it  
Scuola Normale Superiore and Istituto Nazionale di Fisica Nucleare, Piazza dei Cavalieri 7, 56126 Pisa, Italy}
\bigskip 

{\large Paul Sorba}\\ 
\medskip 
{\it  
LAPTh, Laboratoire d'Annecy-le-Vieux de Physique Th\'eorique, 
CNRS, Universit\'e de Savoie,   
BP 110, 74941 Annecy-le-Vieux Cedex, France}
\bigskip 
\bigskip 
\bigskip 

\end{center}
\begin{abstract} 
\bigskip 

The impact of bound states in Landauer-B\"uttiker scattering 
approach to non-equilibrium quantum transport is investigated. We show that the noise power 
at frequency $\nu$ is sensitive to all bound states with energies $\omega_{\rm b}$ satisfying 
$|\omega_{\rm b}|<\nu$. We derive the exact expression of the bound state contribution 
and compare it to the one produced by the scattering states alone. The theoretical and 
experimental consequences of this result are discussed.

\end{abstract}
\bigskip 
\medskip 
\bigskip 

\vfill
\rightline{LAPTH-039/16}
\rightline{IFUP-TH 06/2016}
\newpage
\pagestyle{plain}
\setcounter{page}{1}


\section{Introduction} 
\medskip 

Current fluctuations represent a fundamental characteristic feature of quantum transport 
in systems away from equilibrium. These fluctuations generate noise, which besides spoiling  
the signal propagation, provides \cite{L-98}-\cite{SB-06} the experimental basis of noise spectroscopy. 
Combined with the recent progress \cite{BDD-05, J-13} in the measurement techniques, such  
spectroscopy gives a deeper insight \cite{B-10, K-14} in the mechanism of quantum transport at the 
microscopic level. 

\begin{figure}[h]
\begin{center}
\begin{picture}(500,110)(0,220)
\includegraphics[scale=0.7]{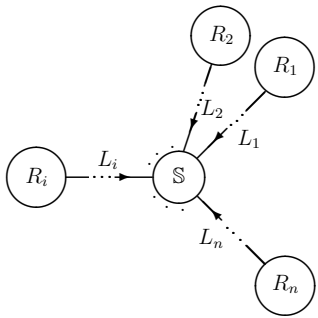}
\end{picture} 
\end{center}
\caption{Multi-terminal system with scattering matrix $\S$ and heat reservoirs $R_i$.} 
\label{fig1}
\end{figure}

Some decades ago Landauer \cite{L-57} proposed a powerful scattering theory framework  
for the derivation of the particle current, which has been further 
developed by B\"uttiker \cite{B-86}. The Landauer-B\"uttiker approach 
goes much beyond the linear response approximation and 
is the core of modern quantum transport theory. It 
has been successfully generalised \cite{A-80}-\cite{SI-86} and applied to the 
computation of the noise power \cite{ML-92}-\cite{BB-00}  
and the full counting statistics \cite{K-87}-\cite{LC-03}.

In this paper we 
address a subtle question in the Landauer-B\"uttiker (LB) scheme, 
namely the impact of bound states on the current fluctuations. 
To be more precise, we consider the multi-terminal system shown in Fig. \ref{fig1}. 
Each semi-infinite lead $L_i$ is attached at infinity to 
a heat reservoir $R_i$ with (inverse) temperature $\beta_i$ and chemical potential $\mu_i$. 
The interaction between the leads is localised 
at their junction and is described by a unitary scattering matrix $\S$. 
The system is away from equilibrium if there are leads which communicate via 
non-vanishing transmission elements of $\S$. Our goal will be to analyse 
the quantum transport in the case when the analytic structure of $\S$ 
implies the existence of bound states with energies $\omega_{\rm b}$. 
In synthesis, we will show that the current fluctuations at frequency $\nu$ are affected by all 
bound states with energies $|\omega_{\rm b}|<\nu$. We will derive in closed and explicit form 
both the scattering and bound state contributions $P_{\rm s}$ and $P_{\rm b}$ to 
the noise power of the system in the LB non-equilibrium steady state. The comparison between 
$P_{\rm s}$ and $P_{\rm b}$ reveals the essential role of bound states. 
We adopt the field theory framework of \cite{MSS-14}-\cite{MSS-16}, 
which allows for a systematic account of all bound states.  

The paper is organised as follows. In section 2 we describe in detail the 
system we are dealing with, focussing on the bound state contribution to 
the spectrum. In section 3 we compute the connected two-point current correlation function 
and extract the noise powers $P_{\rm s}$ and $P_{\rm b}$. In section 4 we 
discuss the general result and illustrate the behavior of the noise power 
for some values of the parameters characterising the system. Section 5 
collects our conclusions. 

\bigskip 

\section{The system} 
\medskip 

We consider systems where both the particle number and the total energy are conserved. 
The particle number conservation implies that the total scattering matrix $\S$ is a direct sum 
of the $m$-body matrices $\{\S^{(m)}\, :\, m=1,2,...\}$. Since the treatment of bound states 
is notoriously a hard task, we focus in what follows on the simplest non-trivial case. 
Namely, we assume that $\S^{(m)}=\II$ for $m>1$ and keep non-trivial only the one-body scattering 
matrix $\S^{(1)} = \S$, which describes the interaction in the junction. This assumption is justified by 
recalling that the idea of freely propagating fermions along the leads 
accounts remarkably well \cite{L-98} for the experimental results \cite{K-96}. 
One possible realisation of this scenario is the Schr\"odinger junction with a point-like defect. 
Our previous analysis \cite{MSS-14}-\cite{MSS-16} shows that this system represents a 
remarkable laboratory for testing general ideas about quantum transport. The study of 
the bound state problem below confirms once again this statement. 

The dynamics along the leads $\{L_i\, :\, i=1,...,n\}$ is fixed by 
the Schr\"odinger equation\footnote{We adopt the natural units $\hbar =c=k_{\rm B}=1$.}  
\begin{equation}
\left (\ri \prt_t +\frac{1}{2m} \prt_x^2\right )\psi (t,x,i) = 0\, ,  \qquad 
x\leq 0,\; i=1,...,n\, , 
\label{eqm1}
\end{equation} 
supplemented by the equal time canonical anticommutation relation 
\begin{equation}
[\psi(t,x_1,i_1)\, ,\, \psi^*(t,x_2,i_2)]_+ = \delta_{i_1i_2}\, \delta(x_1-x_2)\, , 
\label{ccr1}
\end{equation}
where $*$ stands for Hermitian conjugation. 
Since (\ref{ccr1}) reflects the completeness of the eigenstates of the Hamiltonian, it is 
essential in what follows for fixing the bound state contribution to field $\psi$. 

The junction plays physically the role of a point-like defect. 
The associated interaction determines the one-body scattering matrix $\S$, which  
is fixed by requiring that the bulk Hamiltonian $-\prt_x^2$ admits a 
self-adjoint extension in $x=0$. All such extensions are defined \cite{ks-00}-\cite{h-00} 
by the boundary condition 
\begin{equation} 
\lim_{x\to 0^-}\sum_{j=1}^n \left [\lambda (\II-\UU)_{ij} +\ri (\II+\UU)_{ij}\prt_x \right ] \psi (t,x,j) = 0\, , 
\label{bc1} 
\end{equation} 
where $\II$ is the identity matrix, 
$\UU$ is a generic $n\times n$ unitary matrix and $\lambda >0$ is a 
parameter with dimension of mass. Eq. (\ref{bc1}) guaranties unitary time evolution and 
implies \cite{ks-00}-\cite{h-00} the scattering matrix 
\begin{equation} 
\S(k) = 
-\frac{[\lambda (\II - \UU) - k(\II+\UU )]}{[\lambda (\II - \UU) + k(\II+\UU)]} \, ,   
\label{S1}
\end{equation} 
$k$ being the particle momentum. 
It is easily seen that $\S(k)$ is a meromorphic function 
with poles located on the imaginary axis and different from 0. In fact, let $\U$ be 
the unitary matrix diagonalising $\UU$, namely 
\begin{equation}
\UU=\U\, \UU_d\, \U^*\, ,  \qquad 
\UU_d = \diag \left (\e^{2\ri \alpha_1},...,\e^{2\ri \alpha_n}\right )\, ,  
\label{d2}
\end{equation}  
with $-\frac{\pi}{2} \leq \alpha_i \leq \frac{\pi}{2}$. Then 
\begin{equation} 
\S_d(k) \equiv \U^* \S(k) \U = 
\diag \left (\frac{k+\ri \eta_1}{k-\ri \eta_1},..., \frac{k+\ri \eta_n}{k-\ri \eta_n} \right )\, , 
\label{d3}
\end{equation} 
where 
\begin{equation} 
\eta_i = \lambda \tan (\alpha_i)\, . 
\label{d4}
\end{equation} 
The set $\cP_+$ of poles of (\ref{d3}) in upper half-plane 
($0<\alpha_i <\frac{\pi}{2}$) collects the bound states. 
The representation (\ref{d3}) implies that 
the $n$-terminal junction admits at most $n$ different types of bound states. 

Following \cite{BMS-10, BMS-JMP}, we represent the field $\psi$ as a linear combination 
\begin{equation} 
\psi (t,x,i) = \psi_{\rm s} (t,x,i) + \psi_{\rm b} (t,x,i) \, , 
\label{dec1}
\end{equation} 
where $ \psi_{\rm s}$ collects the contribution of the scattering states and 
$\psi_{\rm b}$ that of the bound states. The scattering states read \cite{M-11} 
\begin{equation}
\chi(k;x)=\left [ \e^{-\ri k x}\, \II +\e^{\ri k x}\, \S^*(k)\right ]\, , \quad k\geq 0\, , 
\label{ss}
\end{equation} 
and define the scattering component 
\begin{equation} 
\psis (t,x,i)  = \sum_{j=1}^n \int_{0}^{\infty} \frac{dk}{2\pi } 
\e^{-\ri \omega (k)t}\, \chi_{ij}(k;x) a_j (k) \, , 
\label{psi1} 
\end{equation} 
where $\omega(k) = \frac {k^2}{2m}$ is the dispersion relation and the operators 
$\{a_i(k),\, a^*_i(k)\, :\, k\geq 0,\, i=1,...,n\}$ generate a standard anti-commutation relation algebra 
$\A_{\rm s}$. Using (\ref{psi1}) one gets 
\begin{equation}
[\psis(t,x_1,i_1)\, ,\, \psis^*(t,x_2,i_2)]_+ = 
\delta_{i_1i_2}\, \delta(x_{12})
+\int_{-\infty}^{\infty}\frac{\rd k}{2\pi} \e^{-\ri k\tx_{12}} \S_{i_1i_2}(k)\, , 
\label{ccrs}
\end{equation}
where $x_{12}\equiv x_1-x_2$ and $\tx_{12}\equiv x_1+x_2$. In the 
presence of bound states ($\cP_+\not=\emptyset$), the second term in the right 
hand side of (\ref{ccrs}) does not vanish, which reflects the fact that the scattering 
states (\ref{ss}) are incomplete. Using the Cauchy's integral formula and 
the representation (\ref{d3}), one finds
\begin{equation}
\int_{-\infty}^{\infty}\frac{\rd k}{2\pi} \e^{-\ri k\tx_{12}} \S_{i_1i_2}(k) = 
-2 \sum_{j\in \cP_+} \U_{i_1j} \eta_j \e^{\eta_j \tx_{12}} \U^*_{ji_2}\, . 
\label{cbs}
\end{equation} 
In order to compensate (\ref{cbs}) and satisfy the canonical anti-commutator (\ref{ccr1}), we 
must add new degrees of freedom describing the bound states. For this purpose we 
introduce the algebra $\A_{\rm b}$ generated by the fermion oscillators 
$\{b_i,\, b_i^*\, :\, i \in \cP_+\}$ satisfying 
\begin{equation}
[b_{i_1}\, ,\, b^*_{i_2}]_+ = \delta_{i_1i_2}\, , \qquad  
[b_{i_1}\, ,\, b_{i_2}]_+ = [b^*_{i_1}\, ,\, b^*_{i_2}]_+ = 0\, . 
\label{bosc}
\end{equation}
Assuming that the generators of $\A_{\rm s}$ and $\A_{\rm b}$ anti-commute, we can write 
the bound state component in the form 
\begin{equation}
\psib(t,x,i)=\sum_{j\in \cP_+} \U_{ij} \e^{-\ri t \omega_{\rm b}(\eta_j) +\eta_j x} \sqrt{2\eta_j} b_j 
\label{psib} 
\end{equation}
with $\omega_{\rm b}(\eta)\equiv -\frac{\eta^2}{2m}$. At this point one can directly 
verify that the total field (\ref{dec1}) indeed satisfies (\ref{ccr1}). 

Summarising, each bound state $i\in \cP_+$ gives rise at the level of 
quantum fields to a new degree of freedom described by the oscillator $\{b_i,\, b_i^*\}$. 
The above construction is 
fully determined by the completeness of the Hamiltonian eigenstates and, 
as shown below, allows to evaluate in a systematic way the bound state contribution to the 
particle current and the noise power. 
\bigskip 

\section{Current-current correlation function and noise power with bound states} 
\medskip 

The particle current in our system has the well known form 
\begin{equation}
j(t,x,i)= \frac{\ri }{2m} \left [ \psi^* (\partial_x\psi ) - 
(\partial_x\psi^*)\psi \right ]  (t,x,i) \, . 
\label{curr1}
\end{equation} 
In order to compute correlation functions of $j$, we must fix a representation of 
$\A_{\rm s}$ and $\A_{\rm b}$. Following the work of Landauer 
\cite{L-57} and B\"uttiker \cite{B-86}, for the scattering component $\A_{\rm s}$ 
we take the LB representation, which adapts perfectly to the physical situation 
shown in Fig. \ref{fig1}. Referring for the details to \cite{M-11}, one has in this representation  
\begin{eqnarray} 
\langle a_i^*(k)a_j(p)\rangle &=& 
2\pi \delta (k-p)\delta_{ij}d_i[\omega(k)]\, , 
\label{A4} \\
\langle a_j(p) a_i^*(k)\rangle &=& 
2\pi \delta (k-p)\delta_{ij}\{1-d_i[\omega(k)]\}\, , 
\label{A5}
\end{eqnarray} 
where $d_i(\omega)$ is the Fermi distribution 
\begin{equation}
d_i(\omega)= \frac{1}{1+\e^{\beta_i (\omega -\mu_i)}}\, .
\label{fermi}
\end{equation} 
Since $\psib$ decays exponentially for 
$x\to -\infty$, where the heat reservoirs are located, we adopt for 
$\A_{\rm b}$ the Fock representation, where $b_i$ annihilate the vacuum and 
\begin{equation}
\langle b_i b^*_j \rangle = \delta_{ij}\, , \qquad 
\langle b^*_i b_j \rangle = 0 \, . 
\label{A6}
\end{equation} 
The multi-particle bound states are created by acting with monomials of 
the type $\{b_{i_1}^*\cdots b_{i_m}^*\, :\, m=1,2,... \}$ on the vacuum. 

In this setup the one-point current expectation value equals the 
well known LB result \cite{L-57,B-86}
\begin{equation}
\langle j(t,x,i)\rangle = 
\sum_{j=1}^n \int_0^\infty \frac{\rd \omega}{2\pi} \left [\delta_{ij} -  
|\S_{ij}(\sqrt{2m\omega})|^2\right ] d_j(\omega)\, . 
\label{LB1}
\end{equation}
The bound state contribution shows up in the higher correlation functions, 
starting from the connected two-point function in $L_i$
\begin{equation}
\langle j(t_1,x,i) j(t_2,x,i)\rangle^{\rm conn} \equiv  
\langle j(t_1,x,i) j(t_2,x,i)\rangle - \langle j(t_1,x,i)\rangle \langle j(t_2,x,i)\rangle  \, .
\label{currconn}
\end{equation}
Since the Hamiltonian of our system is a time independent self-adjoint operator, 
the energy is conserved and (\ref{currconn}) depends only on the time difference $t_{12}=t_1-t_2$.  
Then the noise power at frequency $\nu$ in $L_i$ is defined \cite{BB-00} by the Fourier transform 
\begin{equation}
P(\nu;x,i) = \int_{-\infty}^\infty \rd t_{12}\, \e^{-\ri \nu t_{12}}\, \langle j(t_1,x,i) j(t_2,x,i)\rangle^{\rm conn} \, . 
\label{np}
\end{equation} 
In order to simplify the analysis of $P(\nu;x,i)$, it is instructive to represent it in the form 
\begin{equation}
P(\nu;x,i) =  P_{\rm s}(\nu;x,i) + P_{\rm b}(\nu;x,i)\, , 
\label{np1}
\end{equation} 
where $P_{\rm s}$ collects the pure scattering contribution and $P_{\rm b}$ vanishes in 
absence of bound states. 
\bigskip 

\subsection{Scattering state contribution}
\medskip 

We concentrate first on scattering component, which represents 
the finite-frequency generalisation of the result of Martin, Landauer \cite{ML-92} and 
B\"uttiker \cite{B-92}. The direct computation using (\ref{A4}, \ref{A5}) leads to 
\begin{equation}
P_{\rm s}(\nu;x,i) = 
\frac{1}{16\pi m} \sum_{j,l=1}^n \int_0^\infty \frac{\rd \omega}{\sqrt{\omega (\nu+\omega)}} 
d_j(\nu+\omega)L^i_{jl}(x;\nu,\omega) [1-d_l(\omega)]\, , 
\label{nps1}
\end{equation}
with 
\begin{equation}
L^i_{jl}(x;\nu,\omega) \equiv |{\overline \chi}_{ij}(\sqrt{2m(\nu+\omega)};x)\, \partial_x \chi_{il}(\sqrt{2m\omega};x)-
\chi_{il}(\sqrt{2m\omega};x)\, \partial_x {\overline \chi}_{ij} (\sqrt{2m(\nu+\omega)};x)|^2\, , \qquad 
\label{nps2}
\end{equation}
where $\chi$ is given by (\ref{ss}) and the bar means complex conjugation. It is instructive 
to consider the zero-frequency limit 
\begin{equation}
P_{\rm s}(i) = \lim_{\nu\to0^+} P_{\rm s}(\nu;x,i)\, ,  
\label{nps3}
\end{equation}
which turns out to be $x$-independent. In the two terminal case $n=2$ the 
limit (\ref{nps3}) is in addition $i$-independent\footnote{Since,  
$j(t,0,1)=-j(t,0,2)$ due to the Kirchhoff rule.} and provides 
a useful check, reproducing the zero-frequency noise 
power derived in \cite{ML-92}-\cite{BB-00}, namely 
\begin{eqnarray}
P_{\rm s} =  \int_0^\infty \frac{\rd \omega}{2\pi}\bigl \{ [d_1(\omega) +d_2(\omega) -d^2_1(\omega) -d^2_2(\omega)] 
\tau^2(\omega) \nonumber \\
+[d_1(\omega) +d_2(\omega) -2d_1(\omega)d_2(\omega)] \tau(\omega)[1-\tau(\omega)]\bigr \}\, ,
\label{nps4}
\end{eqnarray}
where $\tau(\omega) = |\S_{12}(\sqrt {2m\omega})|^2$ is the transmission probability. The $\omega$-integration in 
(\ref{nps1},\ref{nps4}) cannot be performed explicitly for generic values of the heat reservoir parameters. 
However the integrands have no singularities\footnote{In fact $L^i_{jl}(x;\nu,\omega) \sim \omega + O(\omega^2)$ around $\omega=0$.} in the integration domain 
and the numerical computation is straightforward. 
The examples in section 4 illustrate this fact. 
\bigskip 

\subsection{Bound state contribution}
\medskip 

We turn now to the bound state contribution. Using the explicit form (\ref{psib}) of $\psib$, after some 
algebra one gets 
\begin{equation}
P_{\rm b}(\nu;x,i) = \sum_{l\in \cP_+} \sum_{j=1}^n
\frac{\theta(2m\nu -\eta_l^2)}{2m\sqrt{2m\nu -\eta_l^2}}\, |\U_{il}|^2\, \eta_l\, \e^{2x\eta_l} 
d_j(\nu -\eta_l^2/2m) M^i_j(x;\nu,\eta_l)\, , \qquad 
\label{npb1}
\end{equation}
where $\theta$ is the Heaviside step function\footnote{Fixed by $\theta(0)=1/2$ at the discontinuity point.} and 
\begin{equation}
M^i_j(x;\nu,\eta)= 
|\eta \chi_{ij}(\sqrt{2m\nu -\eta^2};x) -\partial_x \chi_{ij}(\sqrt{2m\nu -\eta^2};x)|^2\, . 
\label{npb2}
\end{equation} 
Using (\ref{ss}) and introducing the combination 
\begin{equation}
\kappa_l = 2m\nu-\eta_l^2\, ,
\label{npb3}
\end{equation}
one gets 
\begin{eqnarray}
P_{\rm b}(\nu;x,i) = \sum_{l\in \cP_+} \sum_{j=1}^n
\frac{\theta(\kappa_l)}{2m\sqrt{\kappa_l}}\, |\U_{il}|^2\, \eta_l\, \e^{2x\eta_l} d_j(\kappa_l/2m)  \nonumber \\
\times 
|(\eta_l-\ri \sqrt {\kappa_l})\e^{2\ri x \sqrt {\kappa_l}} \delta_{ji}+
(\eta_l+\ri \sqrt {\kappa_l})\S_{ji}(\sqrt {\kappa_l})|^2 \, . 
\label{npb4}
\end{eqnarray} 

Eq. (\ref{npb4}) gives in exact and explicit form the bound state contribution $P_{\rm b}$ 
to the noise power in the LB non-equilibrium steady state and represents our main result. 
Let us briefly describe the key features of (\ref{npb4}). 
We observe first of all that $P_{\rm b}(\nu;x,i)$ is a nonnegative 
oscillating function of $\nu$ and $x$, which vanishes in the limits 
$\nu \to \infty$ and $x\to -\infty$. Moreover,  
\begin{equation} 
\lim_{\nu\to 0^+} P_{\rm b}(\nu;x,i) = 0\, ,  
\label{npb5}
\end{equation} 
implying that the bound states do not affect the zero-frequency noise. 
Therefore, for detecting a bound state $l\in \cP_+$ one needs frequencies 
$\nu > \eta_l^2/2m =|\omega_{\rm b}(\eta_l)|$. 
Notice also that the potential 
discontinuity in (\ref{npb4}), due to the Heaviside function, is absent because 
\begin{equation}
|(\eta_l-\ri \sqrt {\kappa_l})\e^{2\ri x \sqrt {\kappa_l}} \delta_{ji}+
(\eta_l+\ri \sqrt {\kappa_l})\S_{ji}(\sqrt {\kappa_l})|^2 \sim \kappa_l + O(\kappa_l^2) 
\label{exp}
\end{equation}
around $\kappa_l=0$. 

In the shot noise regime $\beta_1=\cdots =\beta_n = \beta \to \infty$ one finds 
\begin{eqnarray}
\lim_{\beta \to \infty}P_{\rm b}(\nu;x,i) = \sum_{l\in \cP_+} \sum_{j=1}^n
\frac{\theta(\kappa_l)\theta(2m\mu_j -\kappa_l)}{2m\sqrt{\kappa_l}}|\U_{il}|^2 \eta_l\, \e^{2x\eta_l} 
\nonumber \\
\times |(\eta_l-\ri \sqrt {\kappa_l})\e^{2\ri x \sqrt {\kappa_l}}\delta_{ji}+
(\eta_l+\ri \sqrt {\kappa_l})\S_{ji}(\sqrt {\kappa_l})|^2 \, .  
\nonumber \\
\label{shot}  
\end{eqnarray} 
Therefore the bound state $l\in \cP_+$ 
contributes to the shot noise for chemical potentials   
$\mu_j \geq \kappa_l/2m$. 

Let us clarify finally the role of the factor $|\U_{il}|^2$ in (\ref{npb4}). From the unitarity constraint  
\begin{equation}
\sum_{l\in \cP_+} |\U_{il}|^2 \leq \sum_{l=1}^n |\U_{il}|^2 = 1\, , 
\label{weight}
\end{equation} 
one infers that $|\U_{il}|^2 \in [0,1]$. Therefore $|\U_{il}|^2$ can be interpreted as a weight factor, 
which measures the contribution of the bound state $l\in \cP_+$ to the particle current in the lead $L_i$ for 
different choices of the matrix $\UU$ in the boundary condition (\ref{bc1}). 
It is instructive to consider the following two limiting cases. For generic 
$\UU$ one has $|\U_{il}|^2 \not= 0$ for all $l\in \cP_+$, implying that all bound states 
contribute to the current in $L_i$. For diagonal $\UU$ one has instead $|\U_{il}|^2 =\delta_{il}$, showing 
that only the bound state $i\in \cP_+$, if it exists, contributes to the current in the lead $L_i$. 
\medskip 

To summarise, we derived the bound state contribution to the noise power of 
the Schr\"odinger junction, showing that the frequency $\nu$ is the appropriate 
control parameter for bound state spectroscopy. In fact, by increasing the value of 
$\nu$ one can identify one after the other all bound states of the system. 
We will illustrate this remarkable feature by some examples in the next section. 
\bigskip 

\section{Comparison between $P_{\rm s}$ and $P_{\rm b}$} 
\medskip 

\begin{figure}[h]
\begin{center}
\begin{picture}(0,140)(110,10) 
\includegraphics[scale=0.8]{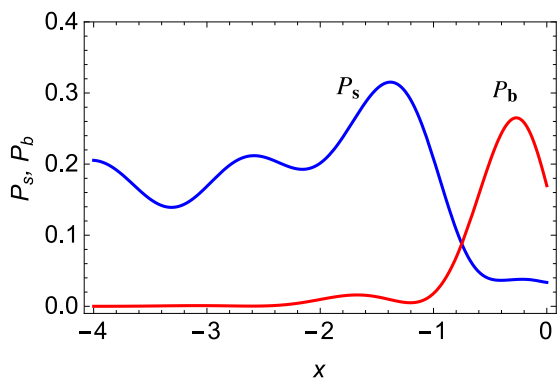}
\end{picture} 
\end{center}
\caption{$x$-dependence of $P_{\rm b}$ and $P_{\rm s}$ at $\nu=3$.} 
\label{fig2}
\end{figure}

We describe now the impact of bound states on the noise power. 
From the unitarity of $\S$ one infers that 
\begin{equation}
|(\eta_l-\ri \sqrt {\kappa_l})\e^{2\ri x \sqrt {\kappa_l}} \delta_{ji}+
(\eta_l+\ri \sqrt {\kappa_l})\S_{ji}(\sqrt {\kappa_l})|^2
\leq 4 m \nu \, . 
\label{com1}
\end{equation}
Inserting (\ref{com1}) in (\ref{npb4}) an using the unitarity of $\U$ one finds the upper bound
\begin{equation}
P_{\rm b}(\nu;x,i) \leq 2\nu \sum_{l\in \cP_+} \sum_{j=1}^n
\frac{\theta(\kappa_l)}{\sqrt{\kappa_l}}\, \eta_l\, \e^{2x\eta_l} d_j(\kappa_l/2m)\, . 
\label{com2}
\end{equation} 
The estimate (\ref{com2}) holds for any lead $L_i$ and shows that the effect 
of the bound states decays exponentially with the distance from the junction. This is 
a simple consequence of the form of the bound state wave functions manifest 
in (\ref{psib}). It turns out that close to the junction the bound state 
noise power $P_{\rm b}$ can dominate the scattering contribution $P_{\rm s}$. 
In order to illustrate this feature we consider the two-lead junction ($n=2$). In this case 
the most general scattering matrix is generated by substituting  
\begin{equation} 
\U= \left(\begin{array}{cc}\e^{\ri \varphi}\cos(\theta/2)
& \sin(\theta/2)\\ 
-\sin(\theta/2) 
& \e^{-\ri \varphi}\cos(\theta/2)  \\ \end{array} \right)\, ,   
\label{u}
\end{equation}
in (\ref{d3}) and has the form 
\begin{equation} 
\S(k)= \left(\begin{array}{cc}\frac{k^2 + \ri k (\eta_1-\eta_2)\cos(\theta)+\eta_1 \eta_2}{(k-\ri \eta_1)(k-\ri \eta_2)} 
& \frac{-\ri \e^{\ri \varphi} k (\eta_1-\eta_2)\sin(\theta)}{(k-\ri \eta_1)(k-\ri \eta_2)}\\ 
\frac{-\ri \e^{-\ri \varphi} k (\eta_1-\eta_2)\sin(\theta)}{(k-\ri \eta_1)(k-\ri \eta_2)} 
& \frac{k^2 - \ri k (\eta_1-\eta_2)\cos(\theta)+\eta_1 \eta_2}{(k-\ri \eta_1)(k-\ri \eta_2)}  \\ \end{array} \right)\, ,   
\label{NN0}
\end{equation}
where $\varphi$ and $\theta$ are arbitrary angles. We assume for illustration that 
two bound states $\eta_1=2$ and $\eta_2=4$ (in units of mass $m$) are present and take  
$\theta=2\pi/3$ and $\varphi = \pi/4$. Finally, fixing the heat reservoir parameters 
$\{\beta_1=1,\, \beta_2=2,\, \mu_1=2,\, \mu_2=3\}$ and setting $m=1$, 
we get the plot in Fig.\ref{fig2}, 
which indeed shows that $P_{\rm b}$ exceeds $P_{\rm s}$ at small $x$. 

\begin{figure}[h]
\begin{center}
\begin{picture}(0,120)(110,20) 
\includegraphics[scale=0.8]{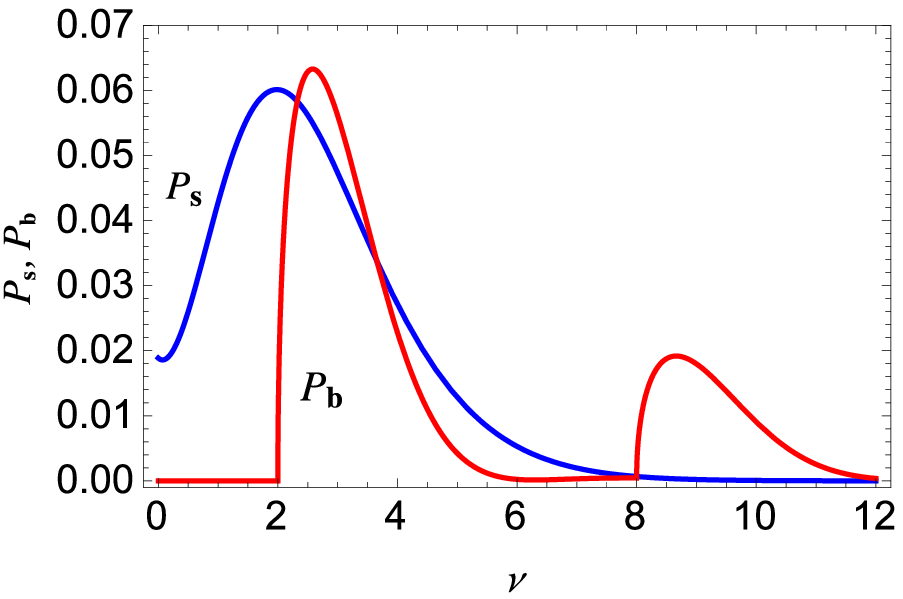}
\end{picture} 
\end{center}
\caption{$\nu$-dependence of $P_{\rm b}$ and $P_{\rm s}$ at $x=-1$.} 
\label{fig3}
\end{figure} 

\begin{figure}[h]
\begin{center}
\begin{picture}(0,120)(110,20) 
\includegraphics[scale=0.8]{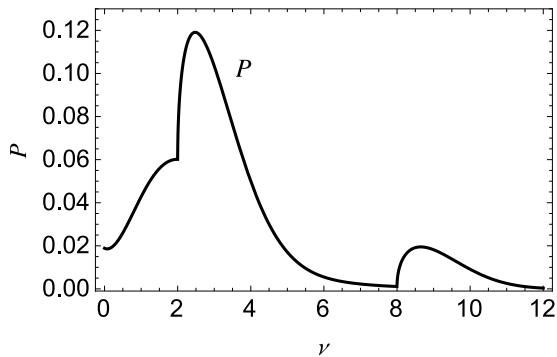}
\end{picture} 
\end{center}
\caption{$\nu$-dependence of $P = P_{\rm b}+P_{\rm s}$ at $x=-1$.} 
\label{fig4}
\end{figure} 

The frequency behavior is much more interesting. Using the same heat bath parameters and 
angles as before, we obtain the plot displayed in Fig. \ref{fig3}. We observe first  
that the bound state contribution is essential and dominating in certain frequency ranges. 
Second, there are discontinuities in the $\nu$-derivative 
of $P_{\rm b}$ at $\nu=\eta_l^2/2m$, which clearly mark the bound states 
and are followed by characteristic resonant-like peaks. 
As already mentioned, this behavior represents an attractive tool for 
bound state spectroscopy. In a finite-frequency experimental setup \cite{FF-12} 
one is usually not able to separate $P_{\rm b}$ from the total noise power 
 $P=P_{\rm s}+P_{\rm b}$. For this reason we report in Fig. \ref{fig4} the plot 
 of $P$, showing that the imprint of the bound states at $\nu=2$ and $\nu=8$ is 
 clearly visible in that case as well. 

Summarising, we have shown that bound states have relevant and specific
impact on both the $x$ and $\nu$-dependence of the finite-frequency noise power and thus
provide interesting experimental signatures.
\bigskip 

\section{Outlook and Conclusions} 
\medskip 

The main scope of this paper was to investigate the role of bound states in the 
Landauer-B\"uttiker scattering approach to non-equilibrium quantum transport. 
For this purpose we considered a simple but nontrivial exactly solvable system, 
namely the Schr\"odinger junction with a point-like defect. In this case the 
total spectrum of the Hamiltonian is fully under control and one can determine 
exactly the scattering and bound state components $P_{\rm s}$ and 
$P_{\rm b}$ of the finite-frequency noise power $P=P_{\rm s}+P_{\rm b}$. 
We established the basic properties of $P_{\rm s}$ and 
$P_{\rm b}$ and showed that the frequency dependence of 
$P$ provides a precise picture of the bound state structure of the system. This 
result finds a direct application to bound state spectroscopy. 

The above framework extends in a straightforward way to the heat current and 
the relative noise power. Moreover, it can be adopted for the study of the bound state 
contribution to the higher current cumulants. It will be interesting in this respect to 
investigate the impact of bound states on the probability distribution, generating 
these cumulants. The extension of the above approach to systems involving Majorana 
bound states, which attract recently some attention \cite{GH-11}-\cite{VGFT-16}, 
is also a challenging open problem. 
\bigskip
\medskip 

\leftline{\bf Acknowledgments:}
\medskip 

M.M. would like to thank the Laboratoire de Physique 
Th\'eorique d'Annecy-le-Vieux for the kind hospitality.

\end{document}